\documentclass{autart}

\usepackage{times,graphicx,latexsym,amsmath,amssymb,color,longtable,wrapfig,wasysym}
\usepackage{datetime,enumitem}

\newtheorem{lemma}{Lemma}

\newcommand{\bv}{\Big\vert}
\begin{document}
\begin{frontmatter}
\title{Quantized Innovations Bayesian Filtering}

\author{Chun-Chia Huang}\ead{chhuang@ucsd.edu}
\author{Robert R.Bitmead}\ead{rrbitmead@ucsd.edu}
\address[ucsd]{Department of Mechanical \&\ Aerospace Engineering,
	University of California, San Diego,
	9500 Gilman Drive, La Jolla CA, 92093-0411, USA.}  

\begin{abstract}
The paper provides simple formulas of Bayesian filtering for the exact recursive computation of state conditional probability density functions given quantized innovations signal measurements of a linear stochastic system. This is a topic of current interest because the innovations signal should be white and therefore efficient in its use of channel capacity and in the design and optimization of the quantizer. Earlier approaches, which we reexamine and characterize here, have relied on assumptions concerning densities or approximations to yield recursive solutions, which include the sign-of-innovations Kalman filter and a Particle filtering technique. Our approach uses the Kalman filter innovations at the transmitter side and provides a point of comparison for the other methods, since it is based on the Bayesian filter. Computational examples are provided.
%
%
\end{abstract}
\end{frontmatter}

\section{Introduction}\label{sec:intro}
Efficient transmission of signals over a limited capacity digital channel requires coding and quantization. Here we examine the recently explored problem of linear state estimation based on quantized innovations signals to reconstruct the conditional probability density function of the predicted and filtered system state given these measurements. We draw on \cite{ARibeiroGB2006,YouXieSunXiao_IMC:2011,R.Sukhavasi&B.Hassibi2013} for algorithmic approaches to this problem, which in turn are based on Bayesian filtering formulations. The Bayesian filter is a recursive approach for the computation of predicted and filtered state conditional probability density functions (pdfs) given a sequence of measurements. It is an approach to nonlinear filtering. Our contribution is to develop for linear stochastic systems a signal transmission system based on the quantized Kalman filter innovations at the transmitter and an associated explicit Bayesian filtering solution at the receiver. We also draw comparisons to other works through the provision of a unifying framework from which to view the development. The advantage of this new approach is that it can benefit from the whiteness of the Kalman filter innovations and its zero-mean and known covariance properties to realize optimal coding and Lloyd-Max optimal quantization \cite{YouAudioCodingTheory:2010}. We also postulate that it has improved error recovery properties because of the reconvergent behavior of the Kalman filter.

\begin{figure}[ht]
  \centering
\includegraphics[width=90mm]{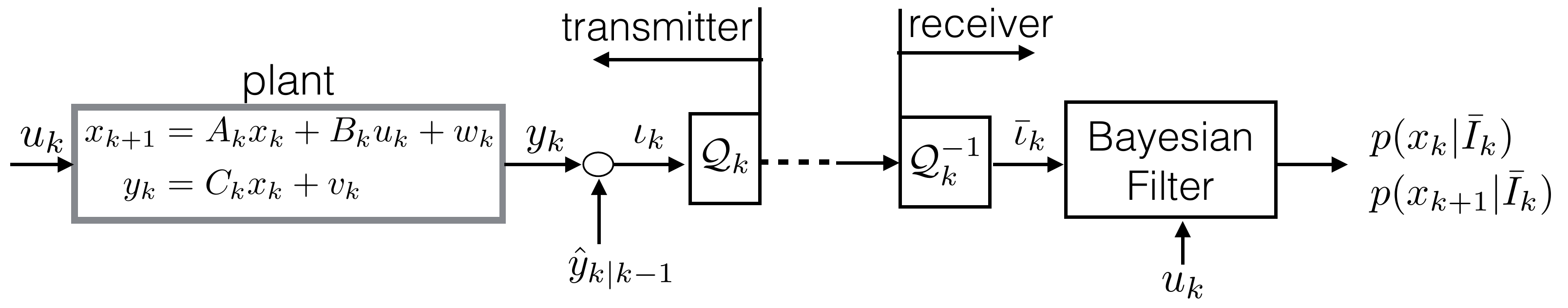}
\caption{Block diagram of quantized innovations system.\label{fig:general_block_diagram}}
\end{figure}

An \textit{innovations} signal, $\iota_k$, is formed by subtracting from a measured signal, $y_k,$ its one-step-ahead prediction, $\hat y_{k|k-1}.$ In signal processing, such calculations are performed in order to whiten a signal prior to transmission over a bandlimited channel and thereby to achieve more efficient use of the channel capacity. In the formal context of Kalman filtering for linear Gaussian systems, the innovations formed from the state prediction is statistically white; a property that we shall apply. The prediction can equally be computed as a reconstruction of a signal available at the receiver, which is achieved at the transmitter by using the computed receiver signal $\bar\iota_k$ and control $u_k$ as the inputs to the predictor. Such a scheme forms the basis of the ITU-T G.726 adaptive differential pulse-coded modulation speech codec standard. The topic of investigation in this paper is the collection of methods using quantized innovations signals for the construction of a state state conditional pdf at the receiver. This is depicted in Figure~\ref{fig:general_block_diagram}, where the reconstruction is performed by a Bayesian filter, which we derive in the sequel. We presume that the input signal, $\{u_k\}$, is available to both ends; a property that occurs in remotely located sensing or in filtering.

\begin{figure}[ht]
  \centering
\includegraphics[width=75mm]{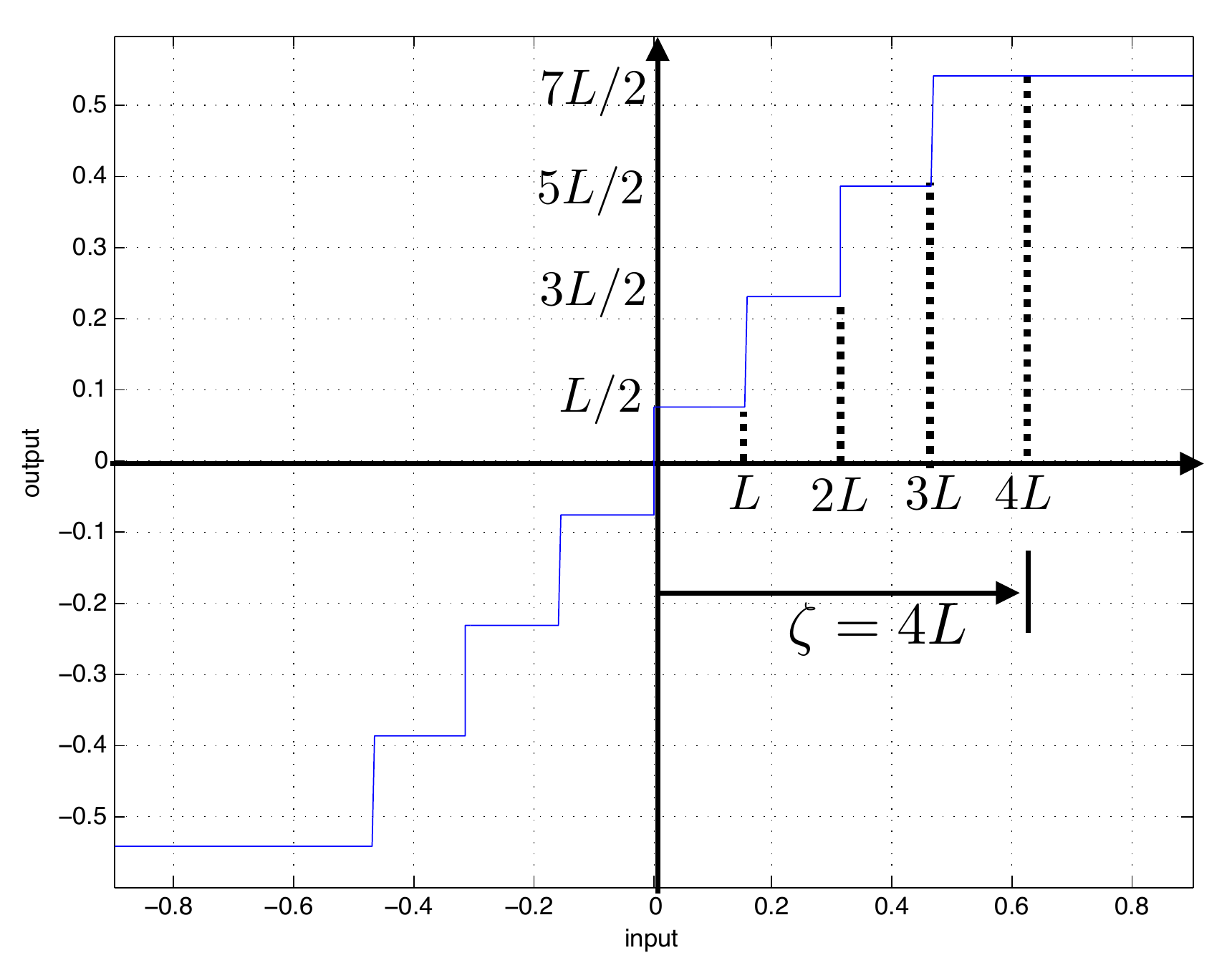}
\caption{Three-bit ($m=8$) uniform quantizer-dequantizer pair input-output relation with $\zeta=0.6222$ and $L=2\zeta/2^{3}=0.1555$. The methods of the paper apply equally to non-symmetric and non-uniform quantizers.\label{fig:quantizer}}
\end{figure}

We use an $m$-level quantizer, $\mathcal Q_k(\cdot),$ consisting of a collection of $m$ intervals $(z_{k,l}, z_{k,u}],$ which form a disjoint covering of the real line and a corresponding rule, $\mathcal Q_k^{-1}(\cdot),$ for dequantizing the received signal into a real number  --- for a $p$-vector channel quantizer, we take a scalar quantizer in each channel. When the innovations signal $\iota_k$ lies within the range $(z_{k,l}, z_{k,u}]$ the output of the quantizer is transmitted as one of $m$ symbols which is then dequantized at the receiver as $\bar \iota_k$ as one of $m$ distinct values, which we take also to be a value within the range $(z_{k,l}, z_{k,u}]$ including for the two edge saturation levels. A $3$-bit uniform quantizer-dequantizer cascade is shown in Figure~\ref{fig:quantizer}. Evidently, $\bar \iota_k = \mathcal Q_k^{-1}\mathcal Q_k(\iota_k)$. The input is the innovations signal, $\iota_k$, and the output is the recovered innovations signal, $\bar \iota_k$. The values $\pm\zeta$ denote the upper and lower saturation limits of the quantizer, which we take for simplicity to be symmetric. Our results apply equally to any well defined quantizer-dequantizer pair without reliance on symmetry or uniformity -- this is important because it admits optimized quantizer designs.

The receiver reconstructs a conditional pdf of the plant state $x_k$ given the received quantized innovations data, $\{\bar\iota_k,\bar\iota_{k-1},\dots,\bar\iota_0\}$. We shall consider three distinct constructions of this measurement signal and corresponding receiver pdfs, distinguished where necessary by superscripts. The detailed algorithms will be presented in Section~\ref{sec:Analysis_tech}.
\begin{description}
\item[Method K:]  A Kalman filter is operated at the transmitter using signals $\{y_k\}$ and $\{u_k\}$ to produce Kalman output prediction $\hat y^\text{K}_{k|k-1},$ and Kalman innovations signal $\epsilon_k$, which is then quantized, transmitted and dequantized at the receiver. 
\begin{align}
\bar\epsilon_k&=\mathcal Q_k^{-1}\mathcal Q_k(\epsilon_k)
=\mathcal Q_k^{-1}\mathcal Q_k(y_k-\hat y_{k|k-1}^\text{K}),\label{eq:Kiota}
\end{align}
The receiver conducts its own processing of this data using a K-Bayesian filter developed in Section~\ref{sec:Analysis_tech}. This is depicted in Figure~\ref{fig:OBF}.
\begin{figure}[ht]
  \centering
\includegraphics[width=88mm]{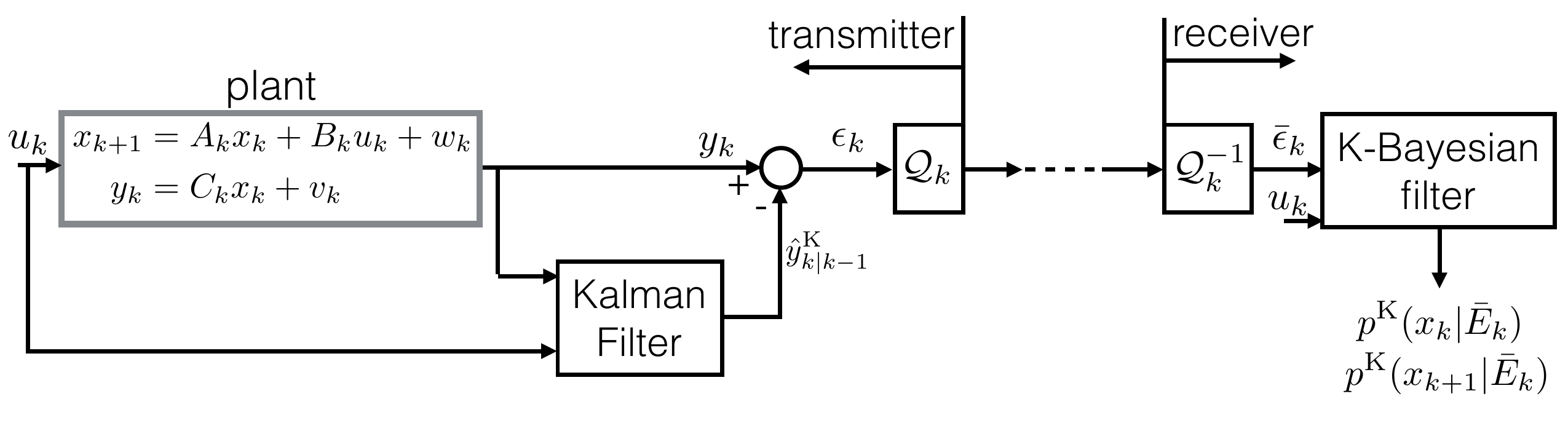}
\caption{Block diagram of Method K showing calculation of the innovations signal, $\epsilon_k$. The K-Bayesian filter is detailed in Section~\ref{sec:Analysis_tech}.\label{fig:OBF}}
\end{figure}

\item[Method R:] The transmitter computes the received signal $\bar\iota_k$ and uses a copy of the receiver's R-Bayesian filter to construct $p^\text{R}(x_k\bv \bar I^\text{R}_{k-1})$, its own version of receiver's conditional state prediction pdf with conditional mean value $\hat x^\text{R}_{k|k-1}$. Then 
\begin{align}\label{eq:Riota}
\bar\iota^\text{R}_k&=\mathcal Q^{-1}_k\mathcal Q_k(y_k-C_k\hat x^\text{R}_{k|k-1}).
\end{align}
This is illustrated in Figure~\ref{fig:Method_R}.
\begin{figure}[ht]
  \centering
\includegraphics[width=88mm]{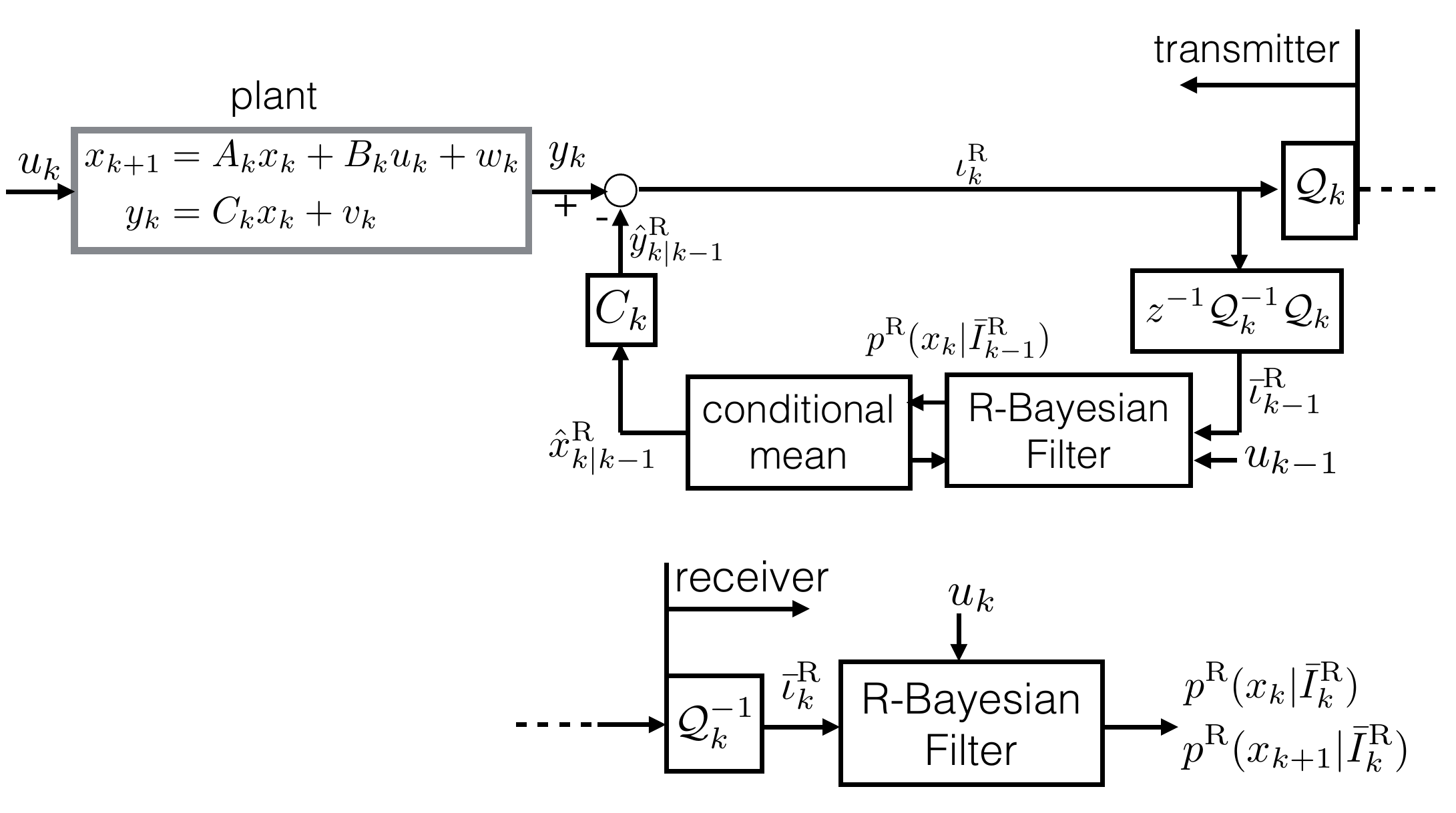}
\caption{Block diagram of Method R showing calculation of the innovations signal, $\iota^\text{R}_k$. The R-Bayesian filter is detailed in Section~\ref{sec:Analysis_tech}.\label{fig:Method_R}}
\end{figure}

\item[Method S:] A modified Kalman filter recursion replaces the R-Bayesian filter of Method R at both transmitter and receiver, yielding synchronized: predicted and filtered state estimates $\hat x^\text{S}_{k|k-1}$ and $\hat x^\text{S}_{k-1|k-1}$, output prediction $\hat y^\text{S}_{k|k-1}=C_k\hat x^\text{S}_{k|k-1}$,  using data $\bar I^\text{S}_{k-1}.$ At the transmitter,
\begin{align}\label{eq:Siota}
\bar\iota^\text{S}_k
&=\mathcal Q^{-1}_k\mathcal Q_k(y_k-C_k\hat x^\text{S}_{k|k-1}).
\end{align}
\end{description}

The Bayesian filter and its progeny Kalman filter are presented in brief detail in Section~\ref{subsec:Bayesian_Kalman} for reference. Section~\ref{sec:Analysis_tech} is then devoted to the explication of the three methods in terms of their algorithms based on the Bayesian and Kalman filter recursions. Computational comparisons are made in Section~\ref{sec:Comparison_examples} and conclusions are then provided. The central contribution of the paper is to recognize the transmitter-side signal model underpinning Method K and its corresponding complete Bayesian filter for the receiver side. The subsequent reinterpretation of Methods~R and S and their comparison are separate contributions enabled by the new Bayesian computation. In the paper we denote: $p(\cdot)$ is a general pdf; $\mathcal N(\mu,\Sigma)$ is the Gaussian pdf with mean value $\mu$ covariance matrix $\Sigma$.

\section{Background and other literature}\label{sec:background}
\subsection{The Bayesian Filter and Kalman Filter} \label{subsec:Bayesian_Kalman}
Suppose we have a state-space system as follows.
\begin{align}
x_{k+1}&=f_k(x_k,w_k),\label{eq:sys_BF}\\
z_k&=h_k(x_k,v_k),\label{eq:out_BF}
\end{align}
where process noise $\{w_k\}$ and measurement noise $\{v_k\}$ are white and of known densities and joint density.  The functions $f_k(\cdot,\cdot)$ and $h_k(\cdot,\cdot)$ are not necessarily linear time-varying system equations. Denoting $Z_k=\{z_0,z_1\ldots z_{k}\}$, the goal of the Bayesian filter is to compute the predicted state conditional probability density function, $p(x_k\bv Z_{k-1}),$ and the filtered pdf, $p(x_k\bv Z_k),$ from an initial density $p(x_0)=p(x_0\bv Z_{-1}).$ 

The Bayesian filter recursion \cite{Simon:2006} is
\begin{align}
p(x_{k} \bv Z_{k-1})&=\int_{x_{k-1}} p(x_k \bv x_{k-1}) p(x_{k-1} \bv Z_{k-1})dx_{k-1},\label{eq:BF_pred}\\
p(x_{k} \bv Z_{k})&=\frac{p(z_k\bv x_k)p(x_k \bv Z_{k-1})}{\int_{x_{k}} p(z_k\bv x_k)p(x_{k}\bv Z_{k-1})dx_{k}}.\label{eq:BF_rec}
\end{align}
\begin{description}
\item[\eqref{eq:BF_pred}] is expressed in terms of $p(x_{k-1}|Z_{k-1})$, the filtered conditional pdf from the other half of the recursion, and $p(x_k|x_{k-1})$ derived from \eqref{eq:sys_BF} using knowledge of function $f$ and the distribution of $w_k$. 
\item[\eqref{eq:BF_rec}] is expressed in terms of $p(x_k|Z_{k-1})$, the predicted conditional pdf from the other half of the recursion, and  $p(z_k|x_{k},Z_{k-1})$ derived from \eqref{eq:out_BF} using knowledge of function $h$ and the distribution of $v_k$. 
\end{description}

For linear Gaussian systems, the Bayesian filter yields the explicit conditional pdf calculation of the Kalman filter. The Kalman filter, however, is much more efficient in that it propagates the Gaussian conditional densities via conditional means and conditional covariances only.
For the linear system,
\begin{align}
x_{k+1}&=A_kx_k+B_ku_k+w_k,\label{eq:statedyn}\\
y_k&=C_kx_k+v_k,\label{eq:output}
\end{align}
with noises $\{w_k\}\sim\mathcal N(0,Q_k)$, $\{v_k\}\sim\mathcal N(0,R_k)$ and  $x_0\sim\mathcal N(\hat x_{0|-1},\Sigma_{0|-1})$ all mutually independent, the Kalman filter recursion is
\begin{align*}
L_{k}&=\Sigma^\text{K}_{k|k-1}C_{k}^T(C_{k} \Sigma^\text{K}_{k|k-1} C_{k}^T+R_{k})^{-1}, \\
\epsilon_{k}&=y_{k}-C_{k}\hat x^\text{K}_{k|k-1}=y_{k}-\hat y^\text{K}_{k|k-1},\\
\hat x_{k|k}^\text{K}&=\hat x_{k|k-1}^\text{K}+L_{k}\epsilon_{k}, \\
\Sigma_{k|k}^\text{K}&=(I-L_{k}C_k)\Sigma_{k|k-1}^\text{K},\\
\Sigma_{k+1|k}^\text{K}&=A_k\Sigma_{k|k}^\text{K}A_k +Q_k, \\
\hat x_{k+1|k}^\text{K}&=A_k\hat x_{k|k}^\text{K}+B_ku_k.
\end{align*}
Therefore, we see the predicted state estimation error signal, $\tilde x^\text{K}_{k+1|k}=x_{k+1}-\hat x^\text{K}_{k+1|k}$, follows the recursion:
\begin{align}
K_k&=A_kL_k,\nonumber\\
\tilde x_{k+1|k}^\text{K}&=(A_k-K_kC_k)\tilde x_{k|k-1}^\text{K}+w_k-K_kv_k ,\label{eq:pred_error}\\
\epsilon_k&=C_k\tilde x_{k|k-1}+v_k.\nonumber
\end{align}
We appeal later to the following properties of the Kalman filter.
\begin{lemma}\label{lem:KFprop}\cite{Anderson&Moore:79}
The Kalman filter innovations is white and
$$\epsilon_k\sim\mathcal N(0,C_k\Sigma^\text{K}_{k|k-1}C^T_k+R_k).$$
\end{lemma}
\begin{lemma}\label{lem:KFconv}\cite{DeystPriceTAC:1968}
Subject to uniform complete detectability of $[A_k,C_k]$, uniform complete stabilizability of $[A_k,Q_k]$, and the existence of $R>0$ such that $R_k\geq R$, the discrete Kalman filter is exponentially asymptotically stable.
\end{lemma}

This section has been devoted to a brief presentation of the Bayesian and Kalman filters based on signal models (\ref{eq:sys_BF}-\ref{eq:out_BF}) and (\ref{eq:statedyn}-\ref{eq:output}) respectively. In the following sections, algorithms will be presented in which the state process, $\{x_k\}$, and measurement sequence, $\{z_k\}=\{\bar\iota_k\}$ or $\{\bar\epsilon_k\}$, will be identified with different signals on the transmitter and receiver sides respectively. In particular, we point out that the signal model for Method~K possesses a state, which we denote $\mathcal Z_k,$ consisting of the concatenation of the plant state and the Kalman filter state at the transmitter. It is this observation which admits the Bayesian filter derivation in this paper.

\subsection{Other literature}
The sign-of-innovations Kalman filter (SOI-KF) was proposed by Ribeiro et al. \cite{ARibeiroGB2006} and yields a precursor to Method~S, applicable in the case of 1-bit quantization of the innovations. Indeed, \cite{ARibeiroGB2006} appears to have been the progenitor for this Bayesian approach to quantized innovations filtering, hence the nomenclature for Method~R derived in Section~\ref{sec:Analysis_tech}. The development of the 1-bit Method~S proceeds from a careful study of the Bayesian filter leading to the statement of Method~R as above but not its precise implementation. We draw on the observations of \cite{ARibeiroGB2006} to make Method~R explicit before proceeding with the analysis of Method~S for the SOI-KF and its extension in the multi-level quantizer Kalman filter (MLQ-KF) of \cite{YouXieSunXiao:WC2008,YouXieSunXiao_IMC:2011}. The advantage of the SOI-KF and MLQ-KF is that a variant of the Kalman filter recursion is computed for approximate conditional means and covariances in place of calculation of the entire pdf. Such reduced calculations suffice because the state predicted conditional pdf is assumed to be Gaussian. This assumption is evaluated (and shown to be somewhat wanting) in the computational examples of Section~\ref{sec:Comparison_examples}.

Sukhavasi and Hassibi in \cite{R.Sukhavasi&B.Hassibi2013} consider the Method K signal model (\ref{eq:statedyn}-\ref{eq:output}) for plant state $x_k$ and demonstrate that the conditional state density given quantized measurements at the receiver is the density of a sum of two independent random variables. The first random variable is the transmitter-side Kalman filter state conditional mean estimate and the second possesses a multivariate truncated Gaussian density involving the complete history of previous quantized measurements. Based on this decomposition, the authors provide a recursive estimation algorithm called Kalman-like particle filter. The decomposition is non-recursive and the algorithm requires approximation for implementation. The authors empirically establish its performance relative to the SOI-KF of \cite{ARibeiroGB2006}.

For the Method~K problem, \cite{ZhanshengDuan&X.Rong2008} observe that, while the predicted pdf, $p(x_k| \bar E_{k-1}),$ can be computed from the traditional Bayesian filtering formula \eqref{eq:BF_pred}, the filtered pdf can be computed by the following formula.
\begin{align}
p(x_k\bv \bar E_k)&=\int_{E_k}p(x_k \bv E_k)p(E_k \bv \bar E_k)dE_k, \label{filter_pdf_nonrecursive5}
\end{align}
where $p(x_k| E_k)$ is a Gaussian probability density function and $p(E_k| \bar E_k )$ is a truncated Gaussian function whose dimension increases with time. That is, this is not a recursive formula. The integration operation in \eqref{filter_pdf_nonrecursive5} is replaced by a recursive formula based on a mid-point approximation to the integrals.

\subsection{Contributions of this paper}
The earlier works either do not yield recursive solutions to the state estimation problem or rely on approximation of the densities to achieve recursion. Our approach, Method~K, is both recursive and exact, subject to the accuracy of computation of the integrals of Bayesian filtering. By the same token, Method~K relies on the computation of these integrals, while Method~S is Kalman-filter-like in its computational demands, which are significantly more modest. The properties described in Lemmata~\ref{lem:KFprop} and \ref{lem:KFconv} imply advantages of Method~K.

\noindent\textbf{Property 1}: Since $\{\epsilon_k\}$ and $\{\bar\epsilon_k\}$ are both white sequences, they are well coded from the perspective that there is no predictable component correlated with past or future data.\\
\textbf{Property 2}: Building on the coding property, since the innovations sequence is Gaussian and of a priori known variance $C_k\Sigma^\text{K}_{k|k-1}C_k+R_k$, the quantizer $\mathcal Q_k(\cdot)$ can be designed to optimize performance using, for example, the Lloyd-Max optimal quantization rule \cite{YouAudioCodingTheory:2010}. This yields maximum entropy coding of the transmitted data.\\
\textbf{Property 3}:  Under the conditions of Lemma~\ref{lem:KFconv}, the transmitter-side Kalman filter forgets exponentially its initial conditions. That is, the dependence of the Kalman filter and of the receiver-side Bayesian filter on initial conditions disappears over time. This property has been studied further for nonlinear filters by van~Handel \cite{vanHandelAAP:2009,vanHandelPTRF:2009}. In the context here, this property suggests that, subject to well-posedness of the Kalman filtering signal model at the transmitter, the Method~K receiver should eventually resynchronize after an isolated error in reception. Similarly, the Bayesian filter integral recursion should exhibit some robustness to inaccuracies in the evaluation of the integrals.

It is not immediately evident that Method~R and its descendent Method~S enjoy these same properties.

\section{Analysis of techniques\label{sec:Analysis_tech}}
In this section, we analyze each method under the following common assumptions. 
\begin{enumerate}[label=\textbf{A.\arabic*}]
\item The plant is linear and described by (\ref{eq:statedyn}-\ref{eq:output}).
\item \label{ass:joint_noise_covariance} Process and measurement noises $\{w_k\}\sim\mathcal N(0,Q_k)$ and $\{v_k\}\sim \mathcal N(0,R_k)$ are independent and white.
\item \label{ass:initial_x0} Initial state $x_0\sim\mathcal N(\hat x_{0|-1},\Sigma_{0|-1})$ is independent from $\{w_k\}$ and $\{v_k\}$.
\item \label{ass:initial_m_Co} The receiver knows: $\hat x_{0|-1}$, $\Sigma_{0|-1}$, and $A_k$, $B_k$, $u_k$, $C_k$, $Q_k$, $R_k$, $\mathcal Q_k(\cdot)$, $\mathcal Q^{-1}_k(\cdot)$, for every value of $k\geq 0.$
\end{enumerate}
\subsection{Method K of this paper}\label{subsec:MethodK}
From \eqref{eq:pred_error}, the receiver side of Method K possesses a state-space and measurement description
\begin{align}
\mathcal Z_{k+1}&=\begin{bmatrix}x_{k+1}\\\tilde x_{k+1|k}^K\end{bmatrix}=
\begin{bmatrix}A_k&0\\0&A_k-K_kC_k\end{bmatrix}
\begin{bmatrix}x_k\\\tilde x_{k|k-1}^K\end{bmatrix}+\nonumber\\
&\hskip 25mm\begin{bmatrix}B_k\\0\end{bmatrix}u_k+
\begin{bmatrix}I&0\\I&-K_k\end{bmatrix}\begin{bmatrix}w_k\\v_k\end{bmatrix},\label{eq:sys}\\
\bar\epsilon_k&=\mathcal Q_k^{-1}\mathcal Q_k\left(\begin{bmatrix}0&C_k\end{bmatrix}\mathcal Z_k+\begin{bmatrix}0&I\end{bmatrix}\begin{bmatrix}w_k\\v_k\end{bmatrix}\right)\label{eq:inn}
\end{align}
By comparison to (\ref{eq:sys_BF}-\ref{eq:out_BF}), this immediately invites the application of the Bayesian filter (\ref{eq:BF_pred}-\ref{eq:BF_rec}) with the substitutions $\mathcal Z_k\mapsto x_k,$ $\bar\epsilon_k\mapsto z_k,$ and $\bar E_k\mapsto Z_k.$
\begin{align} 
p^\text{K}(\mathcal Z_{k}\bv\bar E_{k-1})&=\int_{\mathcal Z_{k-1}}{p(\mathcal Z_{k}\bv\mathcal Z_{k-1})p^\text{K}(\mathcal Z_{k-1}\bv\bar E_{k-1})d\mathcal Z_{k-1}},\label{eq:new_pred}\\
p^\text{K}(\mathcal Z_{k} \bv\bar E_{k})&=\frac{p(\bar\epsilon_{k}\bv\mathcal Z_{k})
p^\text{K}(\mathcal Z_{k}\bv\bar E_{k-1})}{\int_{\mathcal Z_k}{p(\bar\epsilon_k\bv\mathcal Z_k)p^\text{K}(\mathcal Z_k\bv\bar E_{k-1})d\mathcal Z_k}},\nonumber\\
&= \frac{p(\bar\epsilon_{k}\bv\mathcal Z_{k})
p^\text{K}(\mathcal Z_{k}\bv\bar E_{k-1})}{p(\bar\epsilon_{k}\bv \bar E_{k-1})},\nonumber\\
&= \frac{p(\bar\epsilon_{k}\bv\mathcal Z_{k})
p^\text{K}(\mathcal Z_{k}\bv\bar E_{k-1})}{p(\bar\epsilon_{k})}.\label{eq:new_filter}
\end{align}

The above equations form a recursive algorithm to compute the predicted pdf and filtered pdf for each time. We show the detail of each term:
\begin{enumerate}
\item In \eqref{eq:new_pred}, 
\begin{align}\label{eq:joint_noise_pdf}
\hskip -5mm p(\mathcal Z_{k} \bv \mathcal Z_{k-1})&=p\left( w_{k-1} =x_{k}-A_{k-1}x_{k-1}-B_{k-1}u_{k-1}\right)   \nonumber\\ 
&\hskip -17mm \times p\Big(K_{k-1}v_{k-1}=A_{k-1}(\tilde x^\text{K}_{k-1|k-2}-x_{k-1})-B_{k-1}u_{k-1}\nonumber\\
&\hskip -10mm -K_{k-1}C_{k-1}\tilde x^\text{K}_{k-1|k-2}-(\tilde x^\text{K}_{k|k-1}-x_{k-1})\Big).
\end{align}
This probability can be computed from the joint Gaussian distribution of $w_{k-1}$ and $v_{k-1}$, corresponding to Assumption~\ref{ass:joint_noise_covariance}.
\item In \eqref{eq:new_filter}, the whiteness of the innovations signal $\{\epsilon_k\},$ and therefore of $\{\bar\epsilon_k\},$ admits 
\begin{align}&p(\bar \epsilon_{k} \bv \bar E_{k-1})=p(\bar \epsilon_{k})=\int_{z_{k,l}}^{z_{k,u}}p(\epsilon_{k})d\epsilon_{k},\nonumber\\
&=\int_{z_{k,l}}^{z_{k,u}} \mathcal N(0,C_k\Sigma_{k|k-1}^\text{K}C_k^T+R_k)d\epsilon_{k},  \label{eq:bar_E}
\end{align} 
where $(z_{k,l},z_{k,u}]$ is a quantization interval at time $k$.
\item The term $p(\bar \epsilon_{k} \bv \mathcal Z_{k})$ can be computed as follows:
\begin{align}
&p(\bar \epsilon_{k} \bv \mathcal Z_{k}) = \int_{z_{k,l}}^{z_{k,u}}p(\epsilon_{k} \bv \mathcal Z_{k})d\epsilon_{k}, \nonumber \\
& = \int_{z_{k,l}}^{z_{k,u}} p(\epsilon_{k}=C_k \tilde x^\text{K}_{k|k-1}+v_{k})d\epsilon_{k},\nonumber \\
&= \int_{z_{k,l}}^{z_{k,u}}  \mathcal N(C_k \tilde x^\text{K}_{k|k-1},R_{k})d\epsilon_{k}. \label{eq:bar_E_Z}
\end{align}
\item The initial probability density function at the receiver
\begin{align}
p(\mathcal Z_0\bv \mathcal Z_{-1} )&= p(\mathcal Z_0) = p\left(  \begin{bmatrix}x_{0} \\ \tilde x_{0|-1}\end{bmatrix}\right),\nonumber \\ 
&= \mathcal N\left(\begin{bmatrix} \hat x_{0|-1}\\0   \end{bmatrix},\begin{bmatrix}\Sigma_{0|-1} &\Sigma_{0|-1}\\ \Sigma_{0|-1}&\Sigma_{0|-1}  \end{bmatrix}  \right).
\end{align}
The second equality corresponds to Assumption~\ref{ass:initial_m_Co} that the transmitter and receiver are synchronized at time zero. 
\end{enumerate}

We summarize the quantized Kalman filter innovations Bayesian filter, Method~K, as follows.
\vskip 5mm
\noindent\begin{tabular}{| l |}
\hline
\textbf{Method K} algorithm at the receiver.\\
Kalman filter at the transmitter\\
\hline 
{\bf Required function:} $p^\text{K}(\mathcal Z_{k-1}\bv\bar E_{k-1}).$\\
1. Compute $p(\mathcal Z_{k}\bv \mathcal Z_{k-1})$ from \eqref{eq:joint_noise_pdf}. \\
2. Compute {\bf predicted pdf} $p^\text{K}(\mathcal Z_{k}\bv \bar E_{k-1})$ from \eqref{eq:new_pred}. \\
3. Receive $\bar \epsilon_{k}=\mathcal Q^{-1}_k\mathcal Q_k\left[y_k-C_k\hat x^\text{K}_{k|k-1}\right]$.\\
4. Compute $p(\bar\epsilon_{k})$ from \eqref{eq:bar_E} and $p(\bar\epsilon_{k}\bv \mathcal Z_{k})$ from \eqref{eq:bar_E_Z}.\\
5. Compute {\bf filtered pdf} $p^\text{K} (\mathcal Z_{k}\bv \bar E_{k})$ from \eqref{eq:new_filter}.\\
\noindent\rule{80mm}{0pt}\\
\hline
\end{tabular}

\subsection{Method R of Ribeiro et al. \cite{ARibeiroGB2006}} \label{subsec:MethodR}
Method R of \cite{ARibeiroGB2006} uses the Bayesian filter (\ref{eq:BF_pred}-\ref{eq:BF_rec}) with measurement sequence $\bar I^\text{R}_{k-1}=\{\bar\iota^\text{R}_{k-1},\dots,\bar\iota^\text{R}_0\}$ followed by a conditional mean calculation.
\begin{align}
p^\text{R}(x_{k} \bv \bar I^\text{R}_{k-1})&=\int_{x_k} p(x_k \bv x_{k-1}) p^\text{R}(x_{k-1} \bv \bar I^\text{R}_{k-1})dx_k,\label{eq:BF_rec1}\\
p^\text{R}(x_{k} \bv \bar I^\text{R}_{k})&=\frac{p(\bar \iota^\text{R}_{k} \bv x_k, \bar I^\text{R}_{k-1})p^\text{R}(x_k \bv \bar I^\text{R}_{k-1})}{\int_{x_{k}} p(\bar \iota^\text{R}_k\bv x_{k},\bar I^\text{R}_{k-1})p^\text{R}(x_{k}\bv \bar I^\text{R}_{k-1})dx_{k}},\label{eq:BFrec2}\\
\hat x_{k|k-1}^{R} &= E[x_k \bv \bar I^\text{R}_{k-1}]=\int_{x_k} x_k p^\text{R}(x_k\bv \bar I^\text{R}_{k-1})dx_k. \label{eq:SOI_mean}
\end{align}
System state equation \eqref{eq:sys_BF} admits the substitution into \eqref{eq:BF_rec1}
\begin{align}\label{eq:pxx}
p(x_k \bv x_{k-1}) = p(w_{k-1}=x_k-A_{k-1}x_{k-1}-B_{k-1}u_{k-1}),
\end{align}
which forms part of \eqref{eq:joint_noise_pdf} in Method~K. Additionally for Method R, from \eqref{eq:Riota} and the quantizer intervals $(z_{k,l}, z_{k,u}]$,
\begin{align}
&p(\bar \iota_k^\text{R} \bv x_{k},\bar I^\text{R}_{k-1})\nonumber\\
&=p\left(\mathcal Q_k^{-1}\mathcal Q_k[C_k (x_{k}-\hat x^\text{R}_{k|k-1})+v_k] \bv x_{k},\bar I^\text{R}_{k-1}\right),\nonumber\\
&=\int_{z_{k,l}}^{z_{k,u}}p \left (C_k (x_{k}-\hat x_{k|k-1}^\text{R})+v_k \bv x_{k},\bar I^\text{R}_{k-1}\right)d\iota^\text{R}_k,\nonumber\\
&=\int_{z_{k,l}}^{z_{k,u}}\mathcal N\left (C_k(x_{k}-\hat x_{k|k-1}^\text{R}),R_k\right )dv_k.\label{eq:trunc}
\end{align}
\begin{tabular}{| l |}
\hline
\textbf{Method R} of Ribeiro et al. \cite{ARibeiroGB2006} algorithm at transmitter\\and receiver\\
\hline 
{\bf Required function:} $p^\text{R}(x_{k-1}\bv\bar I_{k-1}^\text{R})$\\
1. Compute $p(x_k\bv x_{k-1})$ from \eqref{eq:pxx}. \\
2. Compute {\bf predicted pdf} $p^\text{R}(x_k\bv \bar I_{k-1}^\text{R})$ from \eqref{eq:BF_rec1}.\\
3. Compute the conditional mean $\hat x_{k|k-1}^\text{R}$ from \eqref{eq:SOI_mean}.\\
4. Compute at the transmitter\\
\hskip 15mm\mbox{$\bar\iota^\text{R}_k=\mathcal Q^{-1}_k\mathcal Q_k\left[y_k-C_k\hat x^\text{R}_{k|k-1}\right],$}\\
\hskip 4mm and receive $\bar \iota^\text{R}_k$ at the receiver.\\
5. Compute $p^\text{R}(\bar \iota_k^\text{R} \bv x_{k},\bar I^\text{R}_{k-1})$ from \eqref{eq:trunc}.\\
6. Compute {\bf filtered pdf} $p^\text{R}(x_{k} \bv \bar I^\text{R}_{k})$ from \eqref{eq:BFrec2}. \\
\noindent\rule{80mm}{0pt}\\
\hline
\end{tabular}

\subsection{Method S of Ribeiro et al \cite{ARibeiroGB2006} and You et al. \cite{YouXieSunXiao:WC2008}}\label{subsec:MethodS}
In \cite{ARibeiroGB2006}, Ribeiro et al. derive from their Method~R a Kalman-filter-like recursion based on the one-bit quantization (signum) of the associated innovations signal, $\{\iota^\text{R}_k\}$ depicted in Figure~\ref{fig:Method_R}. This is called the sign of innovations Kalman filter (SOI-KF) and the algorithm operates at both the transmitter and receiver. You et al.  \cite{YouXieSunXiao:WC2008} extend these ideas from one-bit quantization of SOI-KF to the multiple-level quantized innovations Kalman filter (MLQ-KF), which we refer to as Method~S. The derivations of both works operate under the assumption that the predicted pdf $p^\text{S}(x_{k+1}| \bar I_{k}^\text{S})$ is Gaussian. 

Denote the normal density function and $\sqrt{\pi}/2\,\text{erfc}(x)$ as
$$\phi(x)=\frac{1}{\sqrt{2\pi}}\exp(-\frac{x^2}{2}),\,\,\,\,\alpha_{z}=\int_{z}^{\infty}\phi(x)dx.$$
Further, denote the $n^\text{th}$ positive level of a symmetric, mid-rise, $N$-level, scalar quantizer by $(z_{k,l}^n,z_{k,u}^n]$. Then the Kalman-filter-like recursion of Method~S/MLQ-KF for $$|\bar\iota_k|\in(z_{k,l}^n,z_{k,u}^n]$$ is 
\begin{align}
\hat x_{k|k-1}^{S}&=A_k \hat x_{k-1|k-1}^{S},\label{eq:MLQKF_xp}\\
\Sigma_{k|k-1}^{S}&=A_k \Sigma_{k-1|k-1}^{S} A_k^T+Q_k,\label{eq:MLQKF_Sp}\\
\hat x_{k|k}^{S}&=\hat x_{k|k-1}^{S}+\text{sgn}(\bar\iota_k^\text{S})\frac{\phi(z_{k,l}^n)-\phi(z_{k,u}^{n})}{\alpha_{z_{k,l}^n}-\alpha_{z_{k,u}^{n}}}, \nonumber \\ 
 &\hskip 15mm \times\frac{\Sigma_{k|k-1}^{S}C_k^T}{\sqrt{C_k\Sigma_{k|k-1}^{S}C_k^T+R_k}},\label{eq:MLQKF_xf}\\
\Sigma_{k|k}^{S}&=\Sigma_{k|k-1}^{S}-\sum_{n=1}^{N}\frac{[\phi(z_{k,l}^n)-\phi(z_{k,u}^{n})]^2}{|\alpha_{z_{k,l}^n}-\alpha_{z_{k,u}^{n}}|},\nonumber\\
&\hskip 15mm\times\frac{\Sigma_{k|k-1}^{S}C_k^T C_k \Sigma_{k|k-1}^\text{S}}{C_k\Sigma_{k|k-1}^{S}C_k^T+R_k}.\label{eq:MLQKF_Sf}
\end{align}
where $\text{sgn}(\cdot)$ is the signum (or sign) function
\begin{align*}
\text{sgn}(\bar\iota_k^\text{S})=\left\{ \begin{array}{ll}+1&\mbox{if }\bar\iota^\text{S}_k\geq 0 \\ -1&\mbox{if }\bar\iota^\text{S}_k<0  \end{array} \right.
\end{align*}
\begin{tabular}{| l |}
\hline
\textbf{Method S} of Ribeiro et al. \cite{ARibeiroGB2006} and You et al. \cite{YouXieSunXiao:WC2008}\\
algorithm at transmitter and receiver\\
\hline 
{\bf Required data:} $\hat x_{k-1|k-1}^\text{S}$, $\Sigma_{k-1|k-1}^\text{S}$\\
1. Compute \textbf{predicted values} $\hat x_{k|k-1}^\text{S}$ and $\Sigma_{k|k-1}^\text{S}$\\
 \hskip 4mm from time-update (\ref{eq:MLQKF_xp}-\ref{eq:MLQKF_Sp}). \\
2. Compute at the transmitter\\
\hskip 15mm\mbox{$\bar\iota^\text{S}_k=\mathcal Q^{-1}_k\mathcal Q_k\left[y_k-C_k\hat x^\text{S}_{k|k-1}\right],$}\\
\hskip 4mm and receive $\bar \iota^\text{S}_k$ at the receiver.\\

3. Compute \textbf{filtered values} $\hat x_{k|k}^\text{S}$ and covariance $\Sigma_{k|k}^\text{S}$\\
 \hskip 4mmfrom measurement update (\ref{eq:MLQKF_xf}-\ref{eq:MLQKF_Sf}). \\
\noindent\rule{80mm}{0pt}\\
\hline
\end{tabular}
\vskip 5mm

\section{Comparison and examples} \label{sec:Comparison_examples}
In this section we present a few simple simulation comparisons between the methods, noting that Method~K yields the precise conditional densities.

\subsection{Examples}
{\bf Case 1: Almost fixed state, Method~K:} The first example has a state which is almost constant and well measured. It is provided to demonstrate that Method~K is sound. The system has scalar state which obeys
\begin{align*}
x_{k+1} &= x_{k}+w_k, \\ 
y_{k} &= x_{k}+v_k,
\end{align*}
with $w_k\sim \mathcal N(0,0.0001),$ $v_k\sim \mathcal N(0,0.00001),$ $x_0\sim\mathcal N(0,.02).$ The 3-bit quantizer is that depicted in Figure~\ref{fig:quantizer}. 

\begin{figure}[ht]
\centering
\includegraphics[width=85mm]{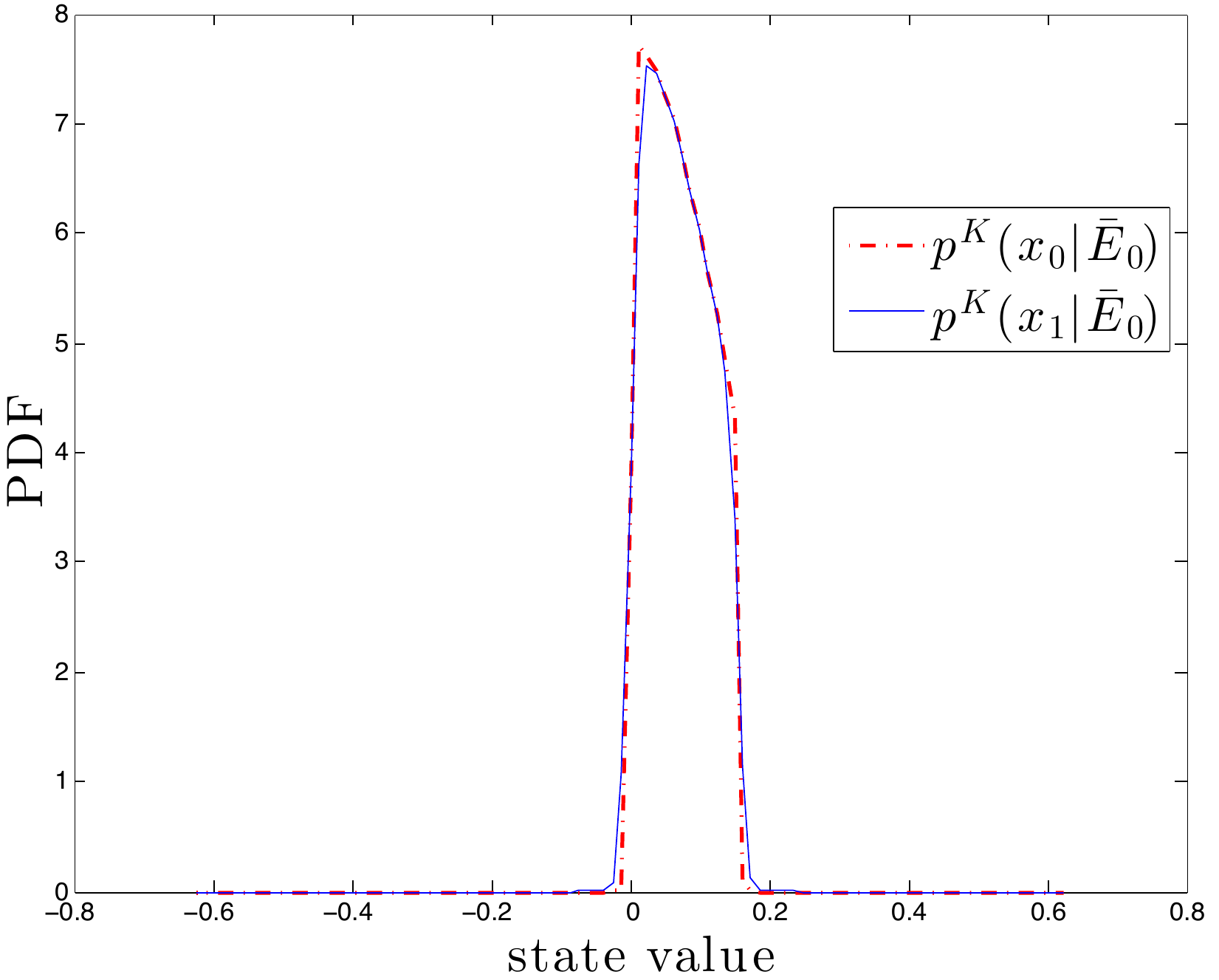}
\caption{Example~1 predicted and filtered pdfs at the receiver computed using Bayesian filtering Method~K with true innovations $ \epsilon_0=0.1160$ and received quantized innovations $\bar\epsilon_0\in (0,0.1555]$ via the 3-bit quantizer in Figure~\ref{fig:quantizer}.\label{fig:fix_point} }
\end{figure}

At the transmitter Kalman filter, the innovations value is $\epsilon_0=0.1160$, leading to a received quantized innovations value $\bar\epsilon_0\in (0,0.1555]$. The reconstructed conditional pdfs are close to truncated Gaussians with slight smoothing at the edges according to the narrow densities of $w_k$ in the predictor and $v_k$ in the filter. Figure~\ref{fig:fix_point} is the result calculated in matlab using 101 sample points for the integrals. These computed pdfs make sense. The next few steps do not alter the figures significantly.

\begin{figure}[ht]
\centering
\includegraphics[width=85mm]{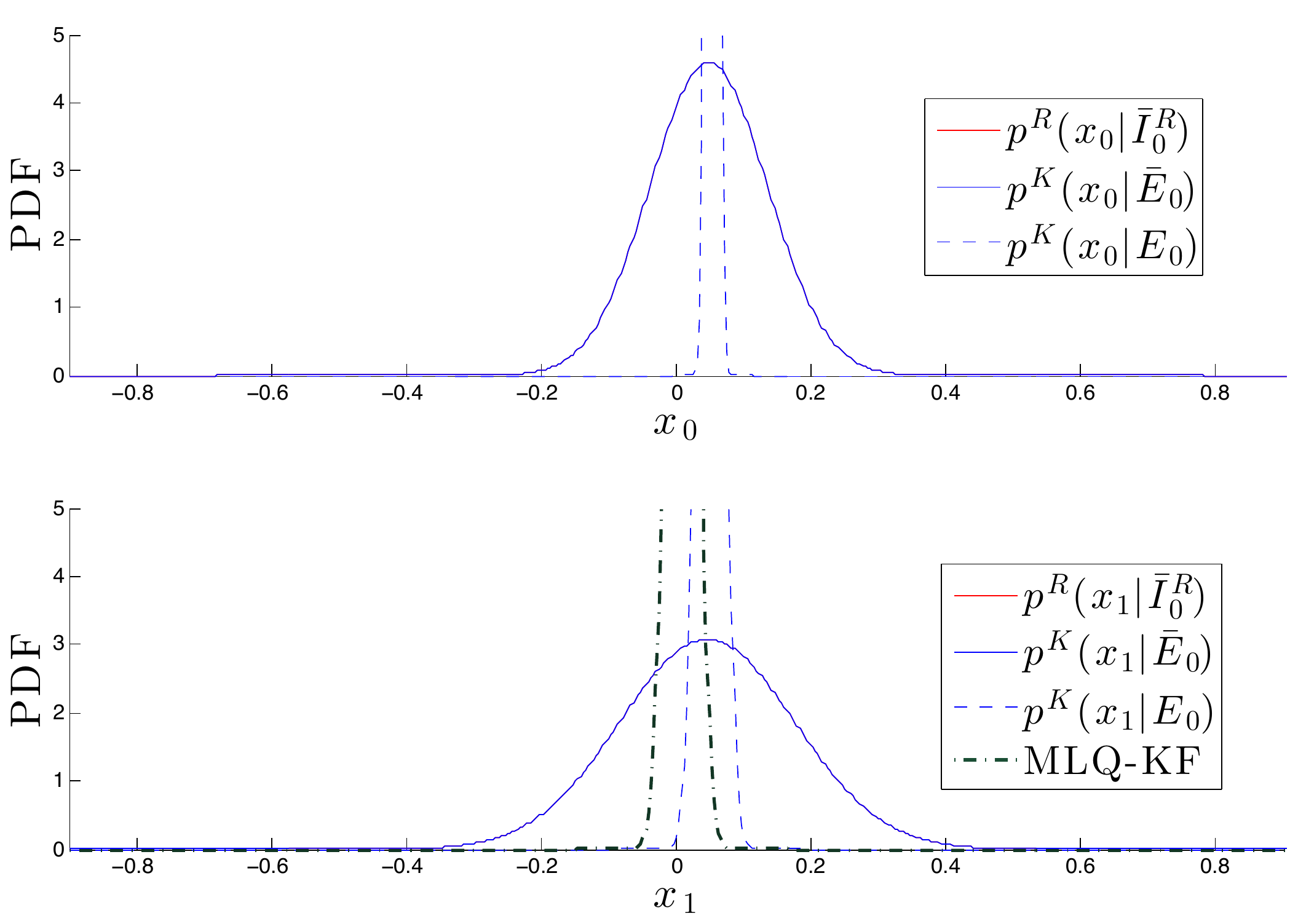}
\caption{Time~0: upper figure filtered state conditional pdfs for Method~K (blue) and R (red) and for the transmitter-side Kalman filter (dashed blue); and lower figure predicted state conditional pdfs for Method~K (blue) and R (red) and Gaussian pdfs for the transmitter Kalman filter (dashed blue) and Method~S/MLQ-KF (dot-dash green). 
\label{fig:t10}}
\end{figure}

\begin{figure}[ht]
\centering
\includegraphics[width=83mm]{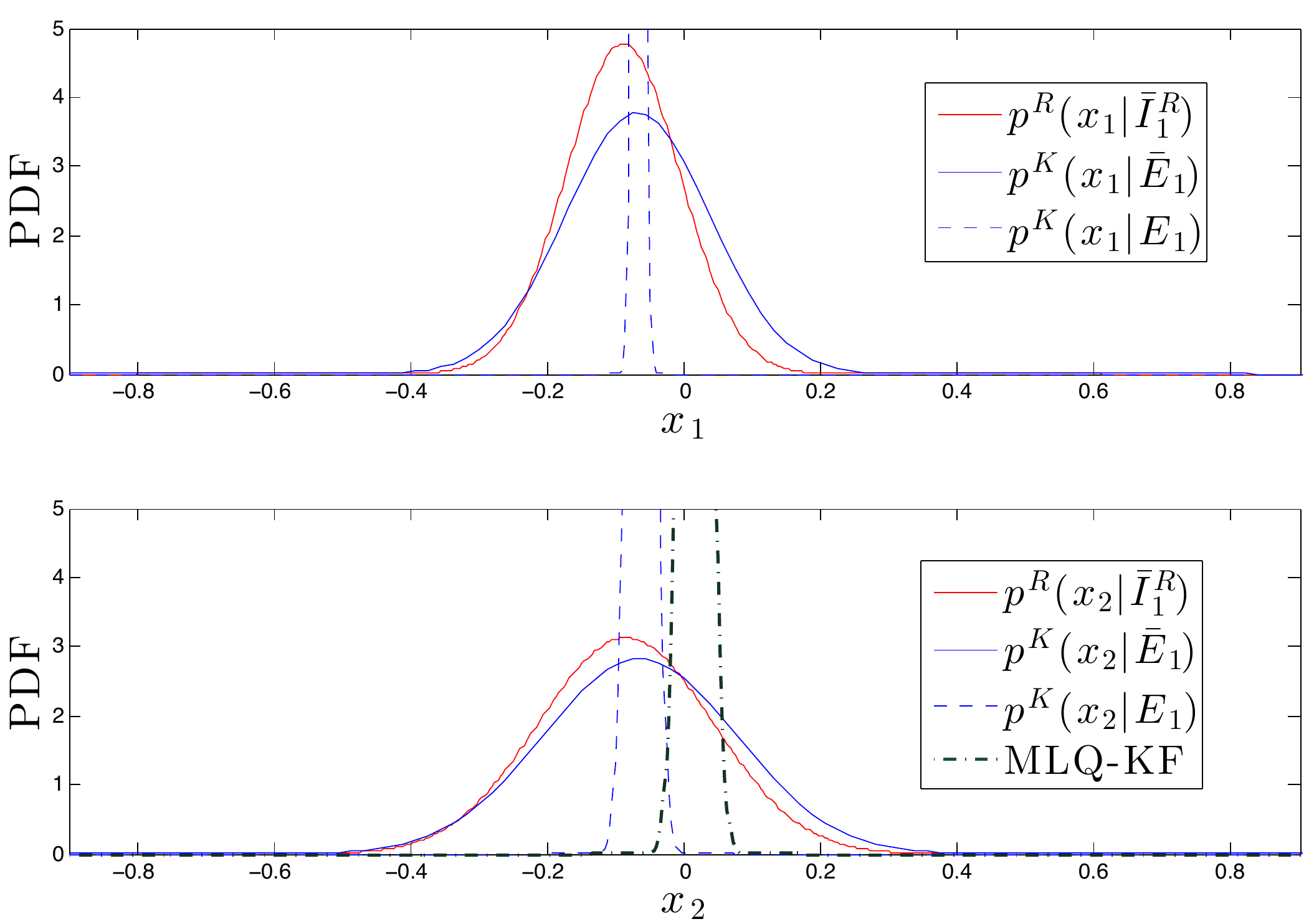}
\caption{Time~1: state conditional pdfs.
\label{fig:t21}}
\end{figure}

\begin{figure}[ht]
\centering
\includegraphics[width=83mm]{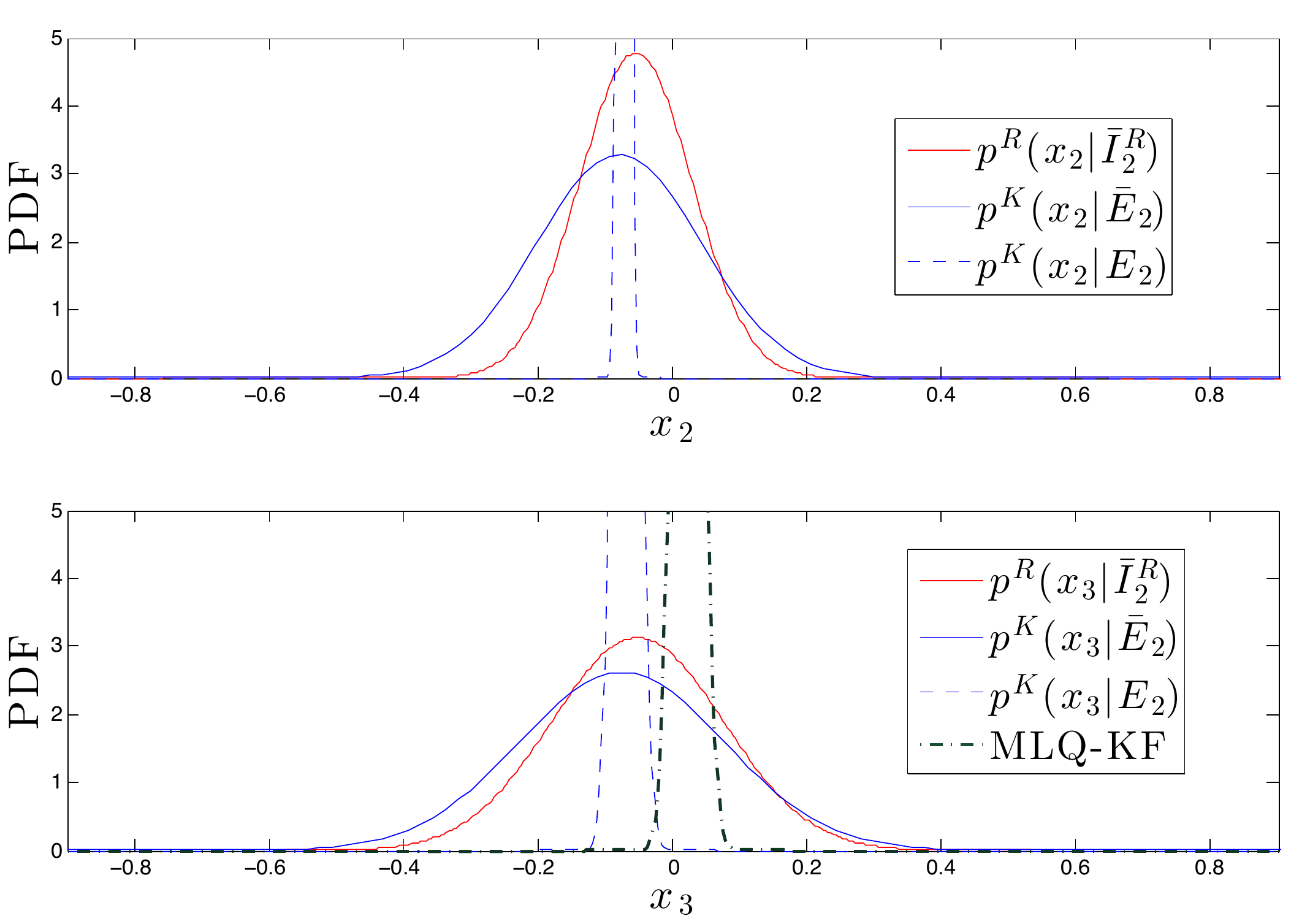}
\caption{Time~2: state conditional pdfs. \label{fig:t32} }
\end{figure}

\begin{figure}[ht]
\centering
\includegraphics[width=83mm]{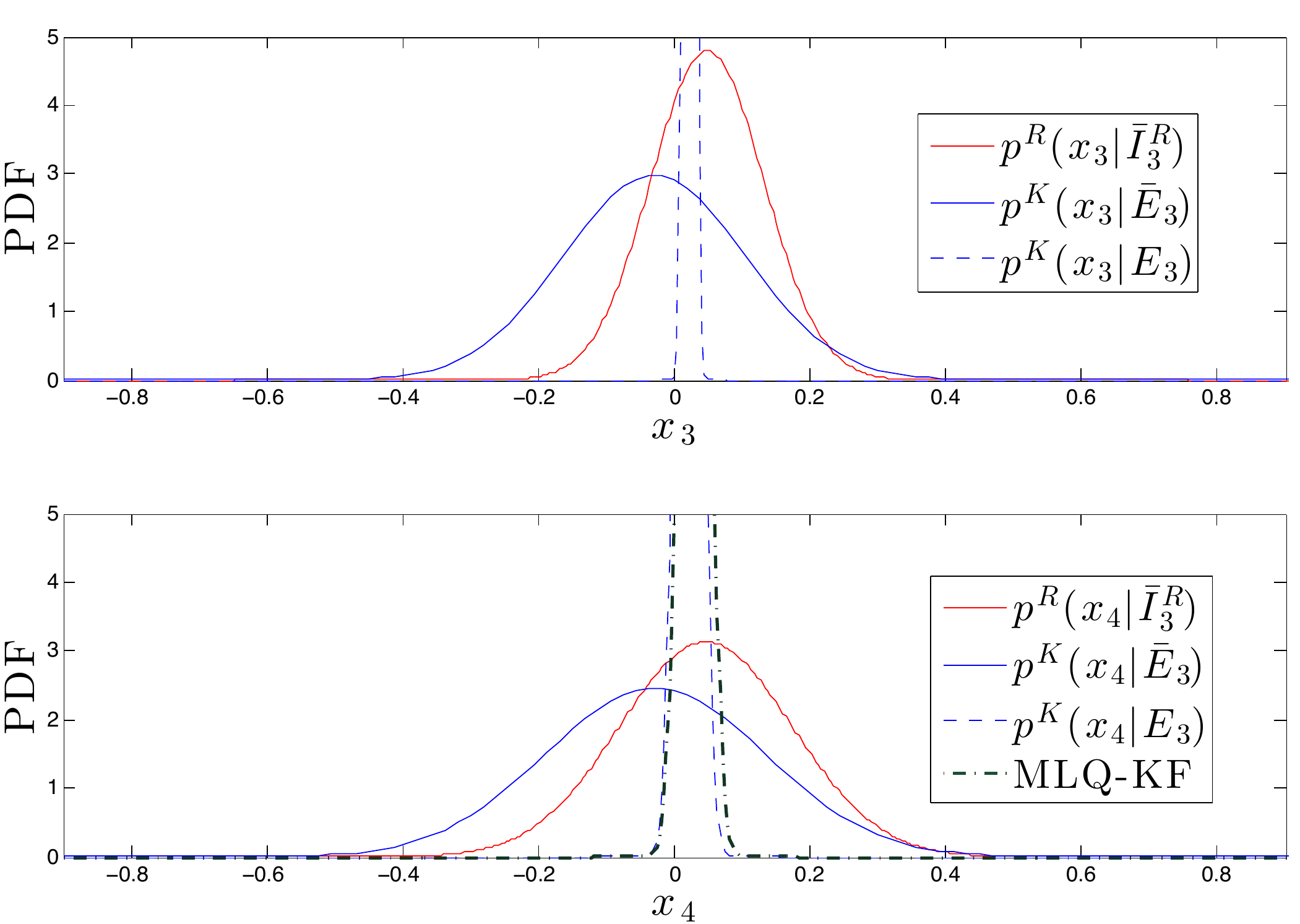}
\caption{Time~3: state conditional pdfs. \label{fig:t43}}
\end{figure}

\begin{figure}[ht]
\centering
\includegraphics[width=83mm]{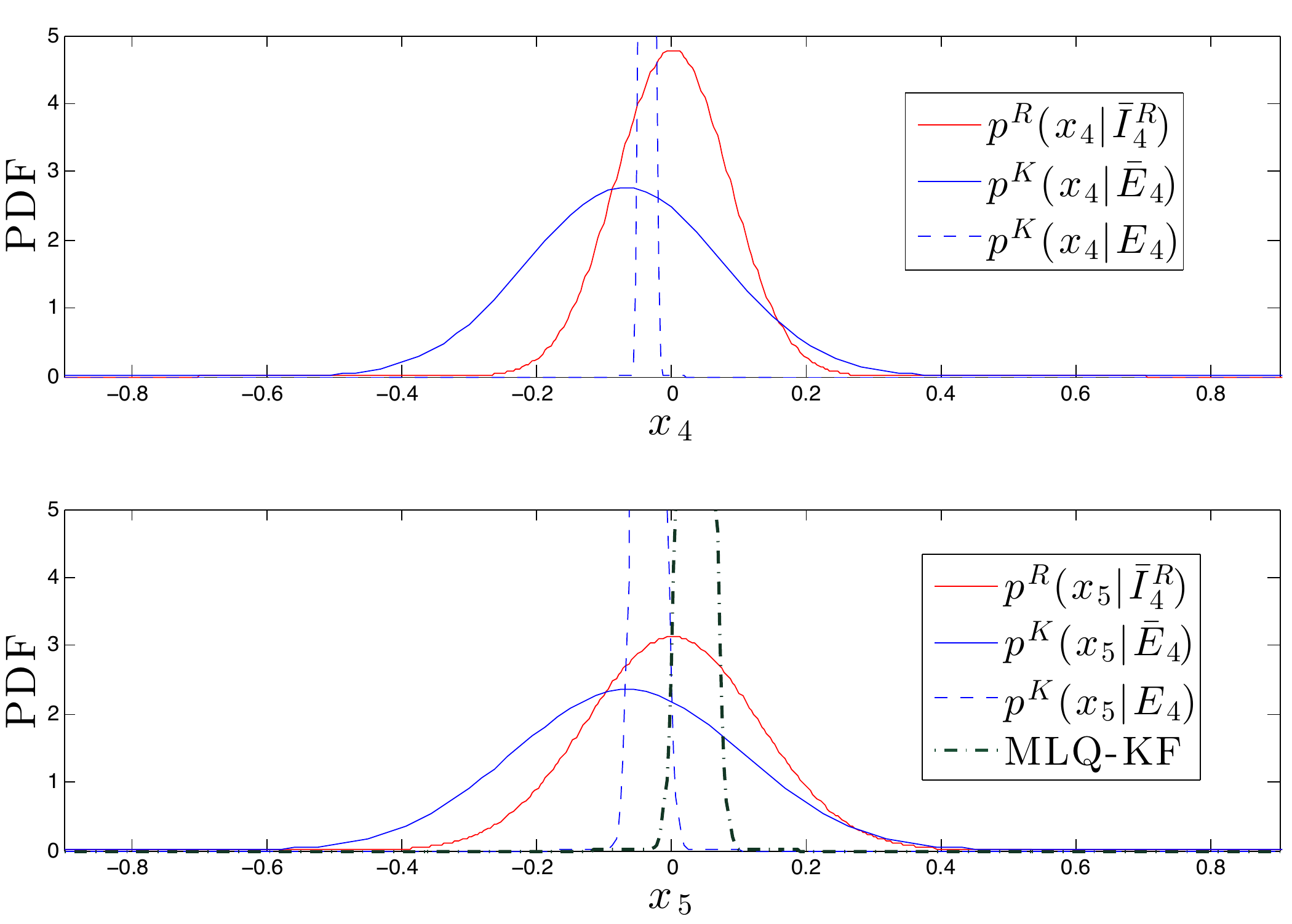}
\caption{Time~4: state conditional pdfs.\label{fig:t54}}
\end{figure}

{\bf Case 2: Comparison of methods:} Consider the system
\begin{align*}
x_{k+1} &= 0.95x_{k}+w_k, \\ 
y_{k} &= x_{k}+v_k. 
\end{align*}
with $w_k\sim \mathcal N(0,0.01),$ $v_k\sim \mathcal N(0,0.01),$ $x_0\sim\mathcal N(0,.02)$ and the same $3$-bit quantizer shown in Figure~\ref{fig:quantizer}. We compare the sequence of predicted and filtered pdfs computed using Method~K and Method~R, and the predicted Gaussian pdf only from Method~S/MLQ-KF using the computed mean and covariance. These are contrasted with the transmitter-side Kalman filter innovations Gaussian pdf. The results are displayed in Figures~\ref{fig:t10}-\ref{fig:t54}.

The signals are are follows.\\
Time~0:
\begin{align*}
x_0&=-0.0319,\; v_0=0.1117,\\
w_0&=-0.1089,\; x_1=-0.1392,\\
\hat x^\text{K}_{0|-1}&=\hat x^\text{R}_{0|-1}=\hat x^\text{S}_{0|-1}=0,\\
\epsilon_0&=\iota_0^\text{R}=\iota^\text{S}_0=0.0798,\\
\bar\epsilon_0&=\bar\iota_0^\text{R}=\bar\iota^{S}_0\in(0,0.1555],\\
\hat x_{1|0}^\text{S}&=0.0085, \;\Sigma_{1|0}^\text{S}=0.0181.
\end{align*}
Time~1:
\begin{align*}
x_1&=-0.1392,\;v_1=0.0033,\\
w_1&=0.0553,\;x_2=-0.0770,\\
\epsilon_1&=-0.1898,\; \iota_{1}^\text{R}= -0.1853,\\ 
\bar\epsilon_1&=\bar\iota^\text{R}_1\in(-0.3111,-0.1555],\\
\hat x_{2|1}^\text{S}&=0.0161,\; \Sigma_{2|1}^\text{S}=0.0177\\
\hat x_{2|1}^\text{K}&= -0.0630,\;\Sigma_{2|1}^\text{K}=0.0156. 
\end{align*}
Time~2:
\begin{align*}
x_2&=-0.0770,\;v_2=0.1101,\\
w_2&=0.1544,\;x_3=0.0813,\\
\epsilon_2&=-0.0140,\;\iota_{2}^\text{R}= -0.1230,\\
\bar\epsilon_2&=\bar\iota^\text{R}_2\in(-0.1555,0],\\
\hat x_{3|2}^\text{S}&=0.0231,\;\Sigma_{3|2}^\text{S}=0.0176,\\
\hat x_{3|2}^\text{K}&= -0.0679,\;\Sigma_{3|2}^\text{K}=0.0155.
\end{align*}
Time~3:
\begin{align*}
x_3&=0.0813,\;v_3=0.0086,\\
w_3&=-0.1492,\;x_4=-0.0720,\\
\epsilon_3&=0.1492,\; \iota_{3}^\text{R}= 0.0352,\\
\bar\epsilon_3&=\bar\iota^\text{R}_3\in(0,0.1555],\\
\hat x_{4|3}^\text{S}&=0.0297,\;\Sigma_{4|3}^\text{S}=0.0175,\\
\hat x_{4|3}^\text{K}&= 0.0216,\; \Sigma_{4|3}^\text{K}=0.0155.
\end{align*}
Time~4:
\begin{align*}
x_4&=-0.0720,\;v_4=-0.0742,\\
\epsilon_4&=-0.0936,\;\iota_{4}^\text{R}= -0.1180,\\
\bar\epsilon_4&=\bar\iota^\text{R}_4\in(-0.1555,0],\\
\hat x_{5|4}^\text{S}&=0.0360,\;\Sigma_{5|4}^\text{S}=0.0175,\\
\hat x_{5|4}^\text{K}&=-0.0335,\;\Sigma_{5|4}^\text{K}=0.0155.
\end{align*}
This second example demonstrates that the methods are workable. Although the explicit computation of pdfs is numerically demanding compared with the moment recursions. However, we note from this sequence of figures that Methods~K and R yield pdfs which circumscribe the transmitter-side Kalman filter pdfs. So the actual state pdf is feasible under the estimated pdf. This appears not the be the case for MLQ-KF at times. Methods~K and R appear comparable and do indeed produce, with these unsaturated quantized data, densities which are roughly gaussian. Example~1 demonstrates that this is not always true, however. If the quantizer were chosen to be Lloyd-Max optimal for Method~K, then one may conjecture that its reconstructed conditional density of the state is itself optimal in some sense. One might then interpret Method~R as producing an approximation of the conditional pdf from Method~K.

\section{Conclusions}\label{sec:conclusion}
We have developed via Bayesian filtering an exact and recursive approach to the reconstruction from quantized Kalman filer innovations signals of predicted and filtered state conditional pdfs. The key observation was to recognize the role played by the Kalman filter state in describing the transmitter signal model. Comparisons were made to a candidate Bayesian approach underpinning the SOI-KF derivation and to the MLQ-KF, indicating some issues with the latter methods. Some specific advantages of the new method of this paper were identified in terms of coding performance, optimal quantization, and recovery from error.

\bibliographystyle{plain}
\bibliography{bobb}
\end{document}